# ELECTRICAL, THERMAL AND OPTICAL CHARACTERIZATION OF POWER LED ASSEMBLIES

*András Poppe[1,2], Gábor Farkas[2], György Horváth[1]*

[1]Budapest University of Technology, Department of Electron Devices, Hungary

[2]MicReD Ltd, Hungary

## ABSTRACT

*Besides their electrical properties the optical parameters of LEDs also depend on junction temperature. For this reason thermal characterization and thermal management play important role in case of power LEDs, necessitating both physical measurements and simulation tools. The focus of this paper is a combined electrical, thermal and optical characterization of power LED assemblies. In terms of simulation a method for board-level electro-thermal simulation is presented, for measurements a combined thermal and radiometric characterization system of power LEDs and LED assemblies is discussed.*

## 1. INTRODUCTION

As it is well known the emitted light output (photometric and/or radiometric flux) of LEDs strongly depends on the junction temperature (as shown in Figure 1 based on data from Lumileds Luxeon Datasheet DS25).

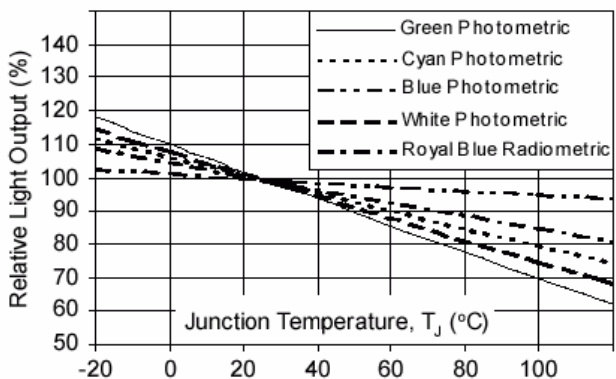

**Figure 1: Temperature dependence of light output for a family of green to blue and white power LEDs**

This is an important issue e.g. in case of LED assemblies where a module realized on a metal-core PCB (MCPCB) contains multiple devices and the uniformity of the light output is important [1]. Such assemblies are often called level 2 devices while single, packaged power LEDs are denoted as level 1.

Prediction of the electrical, thermal and optical properties of level 2 devices requires a *combined electrical-thermal and optical model of level 1* as proposed already in [2]. It also needs proper modeling of the *thermal environment*.

In section 2 we discuss, how LED package thermal models can be obtained directly from measurements, using structure functions. Here we describe a new measurement system which helped us creating a proper LED model. In section 3 we recall the basics of our electro-thermal simulator and its extension to board level simulations utilizing compact thermal models of packages. We conclude the paper with an application example.

## 2. CREATING LED PACKAGE COMPACT THERMAL MODELS

Creating compact thermal models (CTMs) of semiconductor device packages has been discussed in the literature for over a decade. Nowadays the DELPHI approach ([3], [4], [5]) is being widely accepted for creating boundary-condition independent steady-state CTMs, especially for IC packages. A natural extension of these CTMs is towards the transient behavior, in order to obtain dynamic compact thermal models (DCTMs). The methodology for creating such DCTMs as well as extending existing thermal simulators [6] to handle such models was among the achievements of the PROFIT project [7].

When boundary condition independence of the CTM is not required or there exists only a single junction-to-ambient heat-flow path realized by the package, the NID (network identification by deconvolution) method [8] is a viable alternative.

### 2.1. LED package models directly from measurement results

Having a closer look at a typical power LED package in its typical application environment (Figure 2) one may conclude, that the heat generated by the LED chip can basically leave the package along a single path, through the heat-slug, towards the MCPCB substrate.

For steady-state modeling, the package would be perfectly described by a single $R_{thJC}$ junction-to-case thermal resistance value, defined between the LED chip's junction and the heat-slug. This value can be easily identified from thermal transient measurements, using the so-called *double interface method* [9] on the level 1 device.

Another approach is shown in Figure 3 and Figure 4. A level 2 assembly was measured in two setups, denoted

- BC1 – MCPCB directly on cold plate and
- BC2 – MCPCB with a thin plastic sheet inserted between the board and cold plate

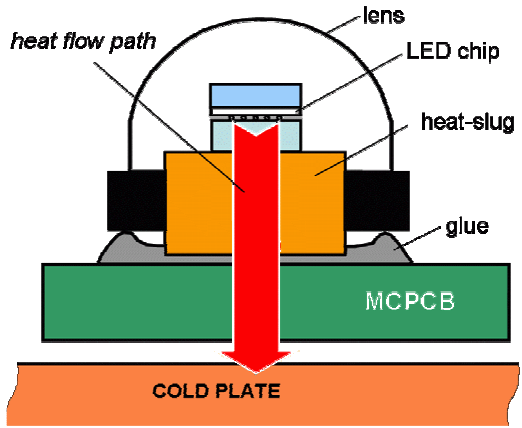

**Figure 2:** Junction-to-ambient heat-flow path in level 2 LED assembly: LED package glued to an MCPCB substrate

In the structure functions $R_{thJC}$ can be read in each function, based on the different thermal conductivity of copper and glue. The inserted sheet causes a change in the structure functions which helps identifying junction-to-board thermal resistance.

If dynamic compact model of the LED package is needed, the single $R_{thJC}$ junction-to-case *thermal resistance* value has to be replaced by a proper model of the *thermal impedance* of the junction-to-case heat-flow path. Structure functions derived from thermal transients are of great help again. The cumulative structure is a map of the thermal capacitances and thermal resistances, smoothly distributed along the junction-to-ambient heat-flow path.

A *step-wise approximation of the cumulative structure function* (Figure 3) can be used to yield a set of lumped thermal resistance and thermal capacitance values for a Cauer-type ladder model of a few stages. (In an earlier NID-based model generation [8], the discretization took place in the time-constant spectrum.)

This method was already successfully used to model stacked-die IC packages [10]. At these packages there can be multiple heat-flow paths, the generated RC ladder model can not be considered as unique model independent of the boundary conditions applied at the different surfaces of the package.

In case of LEDs there is only a single heat-flow path, the RC ladder model of a few stages is perfect dynamic compact thermal model of the LED package.

As the figures suggest at realistic cooling circumstances the structure function based LED package models are *boundary condition independent*, the change of the test environment (in our case: the inserted plastic sheet) has no effect on the portions of the structure function which describe the internal details of the package. In [11] it has been shown that changing the interface at the slug of a level 1 device does not change the early sections either. Thus, the ladder model of Figure 3 describing the heat-flow path until the heat enters the MCPCB is suitable for considering the LED package during board level simulation to characterize single level 2 LED devices (Figure 2) or level 2 assemblies of multiple LEDs (Figure 8). Further details regarding LED package level modeling can be found e.g. in [11].

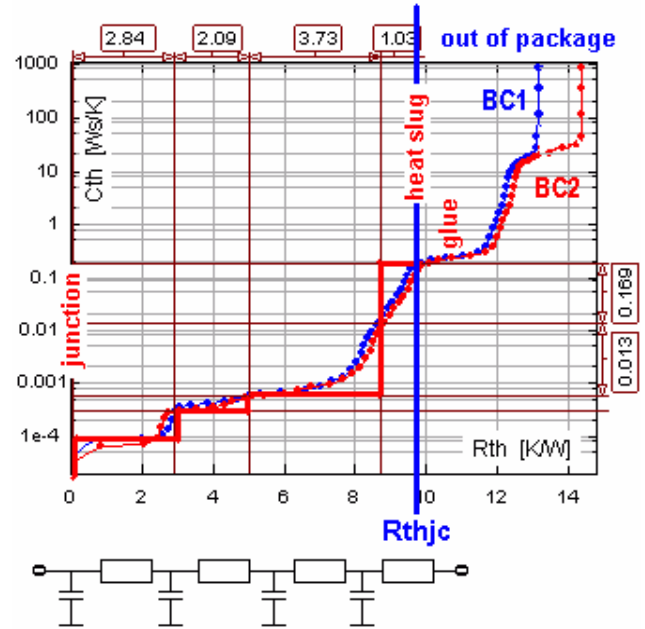

**Figure 3:** Cumulative structure functions: red 1W power LED on MCPCB and its 4-stage package model

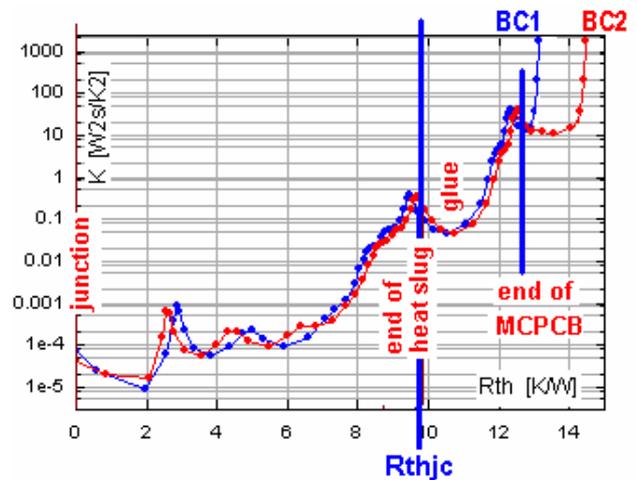

**Figure 4:** Differential structure functions: red 1W power LED on MCPCB

### 2.2. Thermal and radiometric measurement of LEDs

Thermal transient measurements of semiconductor devices are based on the electrical test method [12]. In case of conventional devices the thermal resistance (or thermal impedance for the dynamic case) is calculated from the measured temperature rise and the supplied electrical power. In case of high power LEDs however, this method is not applicable, since about 10-40% of the supplied en-ergy leaves the device in form of *light*. That is why, if LED package models are to be derived directly from thermal measurements, one has to account for the emitted optical power. For this purpose we developed a test environment (Figure 5) which allows a combined thermal and radiometric measurement of power LEDs.

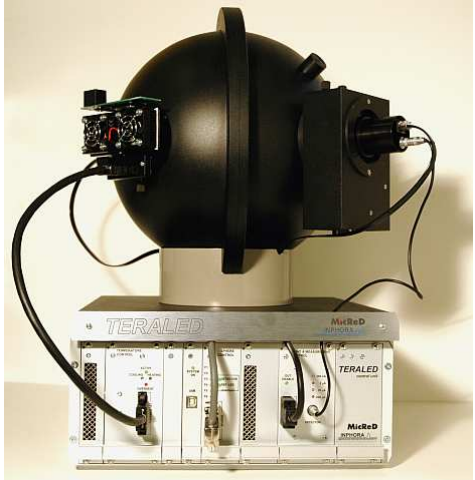

**Figure 5: An automated photometric/radiometric measurement setup with a Peltier-cooled LED fixture, used in connection with the *T3Ster* thermal transient tester**

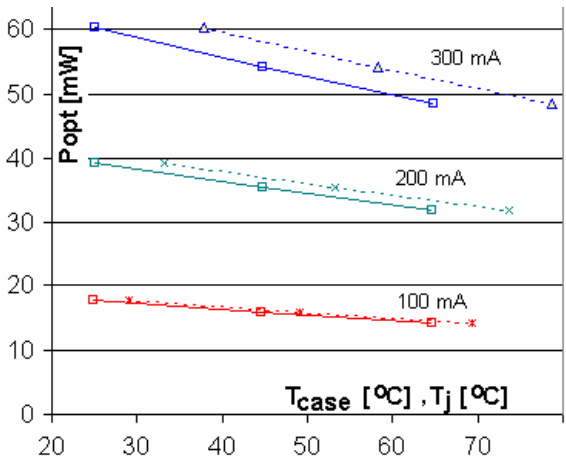

**Figure 6: Emitted optical power of the 1W red power LED as function of the case temperature (solid curves) and junction temperature (dotted curves) at different bias current**

The device under test is mounted to a Peltier-cooled fixture which is attached to an integrating sphere that is in compliance with the applicable standards and recommendations of CIE [13]. The Peltier-cooled fixture stabilizes the LED temperature during the optical measurements and it also serves as a cold-plate for thermal measurements. With a radiometric measurement performed in (thermal and electrical) steady-state of the LED or LED assembly under test, we can measure – among other parameters – the emitted optical power (Figure 6).

Once all optical measurements are performed the DUT LED is switched off and we measure its cooling transient, using the *T3Ster* equipment of MicReD in a usual diode measurement configuration.

The thermal transient test yields the thermal resistance values, so the junction temperatures can also be calculated from the fixture temperatures.

From the raw cooling transient we calculate the thermal impedance curve of the DUT, considering its emitted optical power. This impedance curve is converted into structure functions from which the CTM of the LED package is derived as discussed before.

## 3. ELECTRO-THERMAL SIMULATION ON BOARD LEVEL

### 3.1. Self-consistent electro-thermal simulation with simultaneous simulation

For electro-thermal simulation of electronic circuits containing semiconductor devices we use the *method of simultaneous iteration* [14], [15].

While boundary independence was an important requirement for the *active semiconductor devices* on a substrate (e.g. transistors on the monolithic die or LEDs on an MCPCB) the compact thermal model of the *substrate* itself should reflect the actual conditions at the device interfaces and the connection to the environment. The *boundary condition dependent* substrate model is calculated according to the actual use. The thermal network of the substrate and the devices is solved together with the electrical one, simultaneously.

For connecting the two networks we have electro-thermal semiconductor device models: each device is completed with a thermal node (see Figure 7). The dissipation of the device drives its thermal model network through this thermal node. The electric parameters of the semiconductor device depend on the device temperatures calculated from the complete thermal model. Using the analogy between voltage and resistance, temperature and thermal resistance etc. the response of the coupled electric and thermal network is simulated simultaneously, maintaining a self-consistent solution [16], [17].

### 3.2. Compact thermal model of the substrate

The core of any electro-thermal circuit simulator using simultaneous iteration is how to generate and efficiently handle a boundary condition *dependent* dynamic compact thermal model of the substrate. This thermal model network can be considered as a thermal *N*-port – its ports are terminated by the thermal nodes of the semiconductor devices (Figure 7). This *N*-port model is characterized by *N driving point impedances* describing heat removal from the given semiconductor directly to the ambient and *N*x(*N*-1) *transfer impedances* describing the thermal coupling between any pair of devices on the same substrate.

The NID method uses the time- or frequency domain responses to generate the compact model [8], [18]. These responses are calculated in advance, by using a very fast thermal simulator [19], providing the full set of *N*x*N* time-constant spectra of all the thermal impedances of the substrate. Then the time constant spectra are turned into RC Foster models of a few stages (as the accuracy requires) and these are simulated together with the electrical part in an efficient way [20].

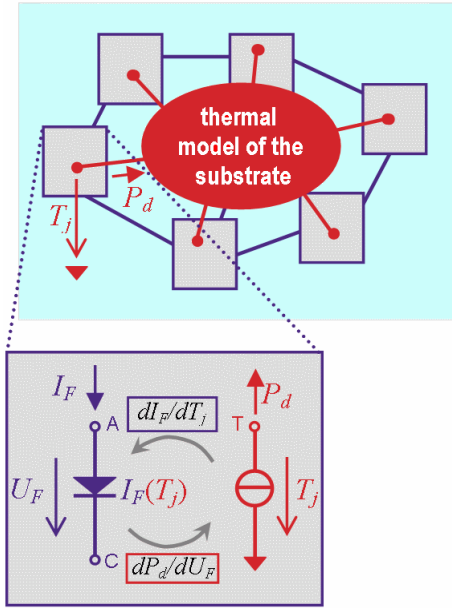

**Figure 7: Sketch of a simple electro-thermal diode model with the *N*-port compact model of the substrate**

### 3.3. Board level extension

The thermal simulator calculates the thermal time-constant spectra for every dissipating shape of the circuit automatically. This algorithm works well for die level IC simulation.

When the temperature dependence of the electric device parameters is not critical we can use a *thermal simulation only* mode. The thermal simulator is already able of using DCTMs of semiconductor device packages. Co-simulating these along with the detailed model of the PWB substrate we can already get device and substrate temperatures [6].

In *electro-thermal simulation* we are interested in thermally influenced electric transient waveforms besides temperature changes. With the recent extension packaged semiconductor devices on any substrate can be represented by their DCTMs in electro-thermal simulation sessions [21]. The *N*-port model of the substrate is calculated in the same way as for a die level problem. In case of packaged devices considered with the DCTMs of their packages the *N*-port compact thermal model of the substrate is extended for the footprint shapes as well.

The DCTMs are inserted between the corresponding footprint nodes of the substrate model and the junction thermal node of the electro-thermal device model and are solved by the electro-thermal circuit simulation engine.

### 4. MULTI-DOMAIN LED MODEL

In case of LEDs one has to subtract the emitted optical power from the supplied electrical power. This is the heating power that is fed into the package compact model:

$$P_{heat} = P_{el} - P_{opt} \quad (1)$$

In our prior study we have shown, that some of the LED devices may have a substantial dissipation on their electrical serial resistance [2]. Thus, the total heating power is distributed between the junction and the series resistance:

$$P_{heat} = P_D - P_{opt} + P_R \quad (2)$$

where $P_D$ denotes the electrical power dissipated on the junction and $P_R$ is the power dissipated on the serial resistance. The model parameters can be easily identified: combined measurements for obtaining $P_{opt}$ were discussed in section 2.2, and the serial resistance of the device can be measured electrically in the same setup.

This serial resistance may be close to the junction, or might reside farther away, this way we can derive so called *hot resistor* or *cold resistor* type thermal models of an LED package. That is, the heat dissipated at the series resistance is fed into the junction-to-footprint heat-flow path either at the junction node, or further away, at a different node of the package model. This is a fact that one has to reflect in the electro-thermal LED model.

### 5. A CASE STUDY

We have investigated the RGB LED module shown in Figure 8. All the three LEDs are in identical packages. In case of the green and blue LEDs of the module even the fine structures close to the junction are similar.

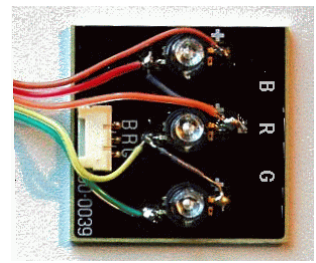

**Figure 8: The investigated LED module**

### 5.1. Measurements

We carried out "plain" thermal transient measurements as well as combined thermal and radiometric ones. The thermal transient measurements were performed in a JEDEC standard still-air environment as well as on cold-plate. In Figure 9 we show the driving point impedance of the green LED measured on cold-plate (Gdriv_CP) and in still-air environment (Gdriv). In the plot one can read at which time and thermal impedance the two heat-flow paths deviate. This measurement result supports our prior statement, that at LED packages one may assume a single path from the junction towards the end of the heat-slug. The plot also shows the transfer impedances in still air (GtoR and GtoB for green driven, red and blue measured). The transfer effect on cold plate can be neglected.

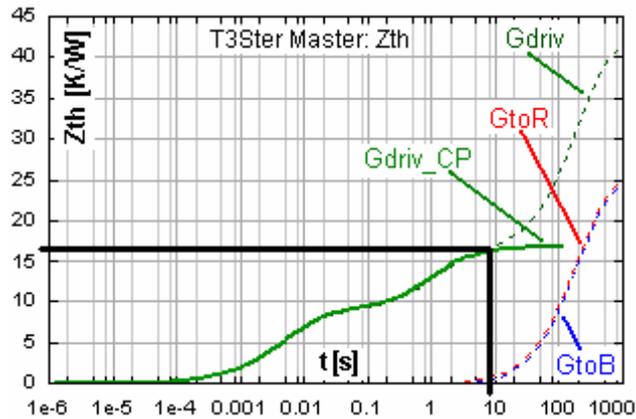

**Figure 9: Thermal impedance curves of the LED module in still air and cold plate setup. Green LED driven, three junctions measured.**

The measurements have also been carried out in the integrating sphere yielding efficiency data for the LEDs. We found that the efficiency of the green LED degrades with increasing cold-plate temperature, similarly to Figure 6.

The DCTM of the LED packages was identified according the procedure described in section 2.1 and was used in board-level simulations of the LED. The electrical part of the LED model was the standard LED model of our electro-thermal circuit simulation engine, parameterized for the actual LED devices.

### 5.2. Simulation

We created the thermal model of the 3 LED module. We used simple Cartesian geometry, circular footprints of the slugs were replaced by a 3x3mm rectangle. The Cauer-type RC ladder model shown in the Spice netlist below has been applied as the DCTM of our LED packages.

```
.SUBCKT LADDER 1 0
C0   1  0   3.644748e-004
R0   1  2   3.178814e+000
C1   2  0   5.871133e-004
R1   2  3   3.125115e+000
C2   3  0   1.036391e-003
R2   3  4   1.605481e+000
C3   4  0   7.580729e-003
R3   4  5   9.286101e-001
C4   5  0   8.746424e-002
R4   5  6   1.113263e+000
.ENDS LADDER
```

Three LEDs have been placed onto an aluminum substrate of 30x30mm$^2$ area and 2.5mm thickness. The thermal model has been obtained from cold-plate measurements. For checking its validity we simulated the LED module in a still-air environment – for which we also had measurement results for later comparison.

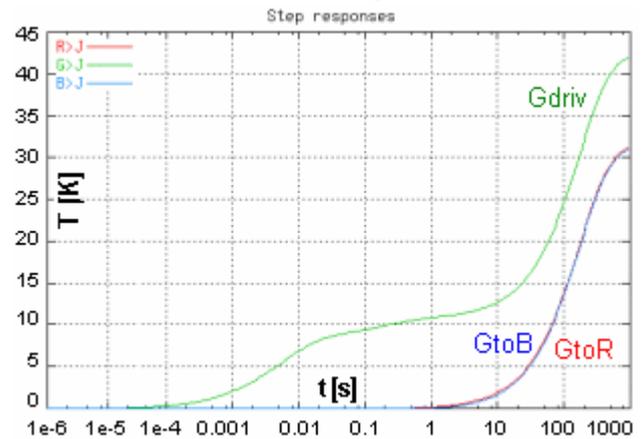

**Figure 10: Simulated thermal impedance plots at the three junctions of the LED module in still air, green LED driven**

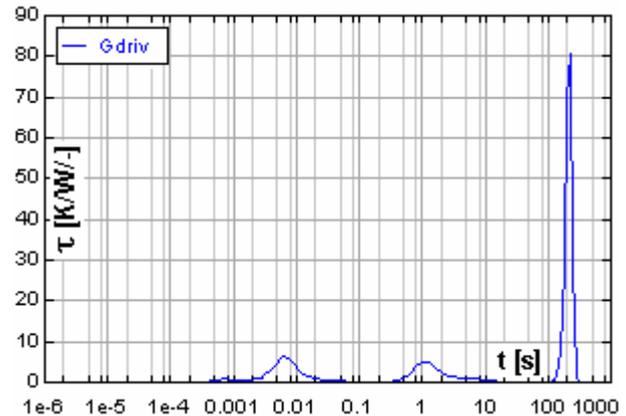

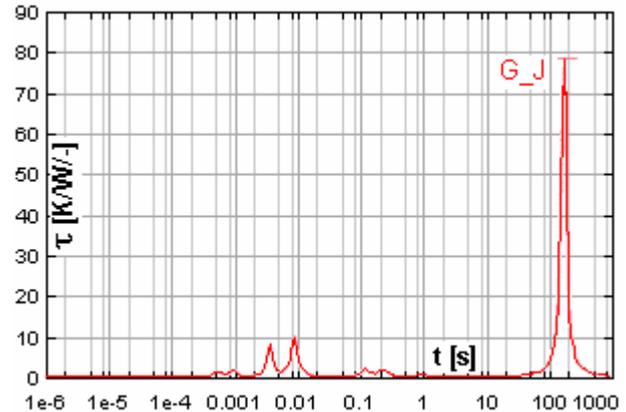

**Figure 11: Measured (above) and simulated (below) time-constant spectrum of the driving point thermal impedance, LED module in still air, green LED driven**

We can see in Figure 10 that the simulated thermal impedance curves agree well with the measured ones (Figure 9). The simulation also properly predicts the thermal delay between the green LED and the other two LEDs: the junction temperature of the blue and red LED starts to increase at about 1s. The time constant spectrum of the driving point impedance (Figure 11) gives also a good image about the match of the simulation results and the measurement results.

Figure 9 also shows that time-constants for structural elements inside the package should fall below 10s. Anything above 10s corresponds to the environment of the LED packages (MCPCB in still-air chamber).

## 6. SUMMARY

In this paper we presented measurement and simulation techniques for the multi-domain characterization of LED devices and LED assemblies. In terms of measurement we have been successfully using a combined thermal and radiometric measurement setup in order to be able to identify the actual thermal power that heats the junction of LED devices. The same measurement setup is also suitable to find the efficiency of the LEDs and to measure its basic electrical parameters – all as functions of temperature.

We have presented a compact model identification method which yield CTMs for LED packages directly from thermal transient measurement results.

Our die level electro-thermal simulation methodology has been extended to board level problems. The CTMs of LED packages have been successfully applied in a board level simulation problem.

## 7. ACKNOWLEDGEMENT

This work was partially supported by AGE-00045/03 TERALED project of the Hungarian Government.

## 8. REFERENCES

[1] T. Treuerniet, V. Lammens: "Thermal management in color variable multi-chip LED modules", *Proceedings of the XXII-nd SEMI-THERM Symposium*, San Jose, CA, USA, 12-16 March 2006, pp. 186-190

[2] G. Farkas et al.: "Electric and thermal transient effects in high power optical devices" *Proceedings of the XX-th SEMI-THERM Symposium*, San Jose, CA, USA, 12-16 March 2004, pp. 168-176

[3] C.J.M. Lasance, D. den Hertog, P. Stehouwer: "Creation and evaluation of compact models for thermal characterisation using dedicated optimisation software", *Proceedings of the XV-th SEMI-THERM Symposium*, 9-11 March 1998, San Diego, USA, pp.189-200

[4] H. Pape, G. Noebauer: "Generation and verification of boundary independent compact thermal models for active components according to the DELPHI/SEED methods",

[5] C.J.M. Lasance: "Recent Progress in Compact thermal models", *Proceedings of the XIX-th SEMI-THERM Symposium*, San Jose, CA, USA, 11-13 March 2003, pp 290-299

[6] M. Rencz, A. Poppe, V. Székely, B. Courtois: "Inclusion of RC compact models of packages into board level thermal simulation tools", *Proceedings of the XVIII-th SEMI-THERM Symposium*, 11-14 March 2002, San Jose, CA, USA, pp. 71-76

[7] C. Lasance: "Final report on the EC-funded thermal project PROFIT", *Proceedings of the 9th THERMINIC Workshop*, 24-26 September 2003, Aix-en-Provance, France, pp. 283-287

[8] V. Székely: "THERMODEL: a tool for compact dynamic thermal model generation", *Proceedings of the 2nd THERMINIC Workshop*, 25-27 September 1996, Budapest, Hungary, pp. 21-26

[9] O. Steffens, P. Szabó, M. Lenz, G. Farkas: "Junction to case characterization methodology for single and multiple chip structures based on thermal transient measurements", *Proceedings of the XXI-st SEMI-THERM Symposium*, San Jose, CA, USA, 14-16 March 2005, pp. 313-321

[10] P. Szabó, M. Rencz, V. Székely, A. Poppe, G. Farkas, B. Courtois: "Thermal characterization and modeling of stacked die packages", *Proceedings of ASME InterPACK '05 – IPACK2005 Coneference*, July 17-22 2005, San Francisco, CA, USA, paper# IPACK2005-73437

[11] J. H. Yu, G. Farkas, Q. van Voorst Vader: "Transient thermal analysis of power LEDs at package and board level", *Proceedings of the 11th THERMINIC Workshop*, 28-30 September 2005, Belgirate, Italy, pp. 244-248

[12] http://www.jedec.org/download/search/jesd51-1.pdf Integrated Circuits Thermal Measurement Method – Electrical Test Method (Single Semiconductor Device) EIA/JEDEC JESD51-1 standard.

[13] http://www.cie.co.at/publ/list.html#standard

[14] V. Székely: "Accurate calculation of device heat dynamics: a special feature of the Trans–Tran circuit analysis program", *Electronics Letters*, Vol 9, no.6, pp. 132-134 (1973)

[15] G. Diegele et al. "Fully coupled Dynamic Electro-Thermal Simulation", IEEE *Transactions on VLSI Systems*, Vol.5, No.3., pp 250-257, 1997

[16] V. Székely et al. "Self-consistent electro-thermal simulation: fundamentals and practice" *Microelectronics Journal*, Vol. 28, pp. 247-262, 1997.

[17] V. Székely et.al.: "Electro-thermal and logi-thermal simulation of VLSI designs", IEEE *Transactions on VLSI Systems*, Vol.5, No.3., pp 258-269, 1997

[18] V. Székely: Identification of RC Networks by Deconvolution: Chances and Limits, *IEEE Transactions on Circuits and Systems-I. Theory and Applications*, Vol. CAS-45, No.3, pp. 244-258, 1998

[19] V. Székely, A. Poppe, M. Rencz, M. Rosental, T. Teszéri: THERMAN: a thermal simulation tool for IC chips, microstructures and PW boards. *Microelectronics Reliability*, Vol. 40, pp. 517-524, 2000

[20] M. Rencz et al.: "Electro-thermal simulation for the prediction of chip operation within the package", *Proceedings of the XIX-th SEMI-THERM Symposium*, San Jose, CA, USA, 11-13 March 2003, pp. 168-175

[21] Gy. Horváth, A. Poppe: "The Sissy electro-thermal simulation system – based on modern software technologies", *Proceedings of the 11th THERMINIC Workshop*, 28-30 September 2005, Belgirate, Italy, pp. 51-54